\documentclass[11pt,a4paper]{article}
\usepackage{longtable}
\usepackage{amsmath}
\usepackage{amssymb}
\usepackage{graphicx}
\usepackage{bm}% bold math
\usepackage[section]{placeins}

\newcommand*{\bea}{\begin{eqnarray}}
\newcommand*{\eea}{\end{eqnarray}}
\newcommand*{\be}{\begin{equation}}
\newcommand*{\ee}{\end{equation}}

\newcommand{\bma}{\begin{pmatrix}}
\newcommand{\ema}{\end{pmatrix}}
\usepackage[square,comma,sort&compress,numbers]{natbib}
\title{Two Higgs Doublet Model with Scalar Mediation via Yukawa Interactions}
\author{Tajdar Mufti \footnote{tajdar.mufti@gmail.com, tajdar.mufti@lums.edu.pk}, Touqeer Zahid \footnote{touqeerzahid1@gmail.com} \\ Lahore University of Management Sciences\\ Opposite Sector U, D.H.A, Lahore Cantt., 54792, Pakistan}
\begin{document}
\maketitle
\begin{abstract}
Scalar sector offers a valuable avenue to explore its implications in the particle physics phenomenology for new physics and several model dependent phenomena. Two Higgs doublet models are among the promising models for such explorations. In the presence of a standard model Higgs, a two Higgs doublet model is studied in which the two Higgs fields are connected with each other via a scalar singlet field. The study is conducted using Dyson Schwinger equations under the Yukawa vertices set at their tree level form up to certain constants. Among the field propagators, the two Higgs propagators are found to be least effected in the parameter space which also reflects in the insignificant beyond the bare mass contributions to the renormalized Higgs masses. Evidence of universality in terms of identical coupling for both Higgs is found in the model. The model is found to be non-trivial in the explored parameter space. There are cutoff effects in the model. Stability in the calculated correlation functions and parameters generally ensues beyond 100 TeV cutoff.
\end{abstract}
\section{Introduction} \label{draft:intro}
The standard model (SM) \cite{Schwartz:2013pla,Barnett:1996yz} has already seen its limitations as it is inadequate for complete understanding of a number of observations and the new physics. Supersymmetry \cite{Martin:1997ns,Aaboud:2017leg}, cosmic inflation \cite{Bezrukov:2013fka,Linde:2005ht,Mukhanov:2005sc,Enckell:2018kkc,Guth:1980zm,Ferreira:2017ynu,Hakim:1984oya,Linde:1993cn}, dark matter \cite{Athron:2017kgt,Lee:2017qve,Bento:2000ah,Bertolami:2016ywc,Munoz:2017ezd,Poulin:2018kap}, baryon asymmetry, and naturalness problem are some of these examples. In order to study new physics, various extensions of the SM are studied \cite{Robens:2019kga}. Scalar sector is no exception, and the discovery of the SM Higgs \cite{pdg,Kane:2018oax,Carena:2002es} serves as a strong motivation to study scalar interactions in presence of the SM Higgs. Furthermore, scalars are among the new particles being looked for \cite{Kling:2020hmi} on the experimental front of particle physics searches. A thorough understanding of scalar interactions will be immensely helpful in identifying the new physics.
\par
Historically, studies in Yang-Mills theory \cite{Yamanaka:2019yek,Bergner:2020mwl}, which is the gauge sector in the SM \cite{Schwartz:2013pla,Barnett:1996yz}, has been particularly useful in understanding QCD-related physics \cite{Suenaga:2019jjv,Bazavov:2020teh,Acharyya:2017uhl}. One may harbor similar expectations from pure scalar sector \cite{Hasenfratz:1988kr,Gliozzi:1997ve,Weber:2000dp}, particularly for richer renormalizable quantum field theory models. Among simpler but yet potent models are the ones with Yukawa interactions which can also create larger renormalizable quartic reactions. Discovery of the Higgs boson presents an opportunity to study the sector in order to explore the masses and correlation functions in renormalizable theories. The Wick-Cutkosky (WC) model, and its variants, are among the widely explored models in scalar sector \cite{Darewych:1998mb,Sauli:2002qa,Efimov:2003hs,Nugaev:2016uqd,Darewych:2009wk}.
\par
It is not known apriori if the new Physics is perturbatively accessible. Hence, non-perturbative approaches \cite{Rivers:1987hi,Ruthe:2008rut} naturally find their place in this arena. The method of Dyson Schwinger Equations (DSEs) \cite{Schwinger:1951ex,Schwinger:1951hq,Swanson:2010pw,Roberts:1994dr,Rivers:1987hi,Ayala:2006sv} is one of the widely used approaches for non-perturbative studies.
\par
This paper addresses a two Higgs doublet model (2HDM) \cite{Gunion:2002zf,Cabrera:2020lmg,Altmannshofer:2020shb} with both Higgs fields preserving the SU(2) symmetry and a real scalar singlet mediating field \footnote{For simplicity. one of the Higgs is termed as \textit{the SM Higgs} with its physical (renormalized) mass fixed at $125.09$ GeV, while the other Higgs is called \textit{the second Higgs} whose physical mass is determined in the parameter space of the theory.}. The theory is essentially the case of two complex doublet scalar families interacting with each other only via a scalar particle. For such a model, there are two peculiar cases. First, it presents an opportunity to observe how a particle with mass around $1$ MeV, which is the case with electron and the lightest quarks, interacts with the SM Higgs under the given vertices for various couplings, technical details are given in the next section. Second, the case of the second Higgs mass in TeVs offers an opportunity to understand the physics for the case of two families of particles of masses at different orders of magnitude interacting with each other via a real scalar mediating field. Despite the simplicity, the model has a rich parameter space. Hence, the model is studied for different bare masses and bare couplings with different cutoff values.
\par
During the entire study, the two vertices are represented by their bare Yukawa couplings, similar to ladder approximation \cite{Roberts:1994dr}, up to certain factors due to renormalization. The details are included in the next section. The method of DSEs \cite{Rivers:1987hi} is used to study the model.
\par
There is a companion paper which encompasses a study of a richer model than the one studied here, using a different approach \cite{Ruthe:2008rut}.
\par
It is assumed that the model is not trivial. The assumption is supported by the fact that Higgs interaction with gauge bosons does not render the model trivial \cite{Maas:2013aia,Maas:2014pba}, despite that $\phi^{4}$ theory \cite{Hasenfratz:1988kr,Gliozzi:1997ve,Weber:2000dp} is found trivial \cite{Jora:2015yga,Aizenman:1981zz,Weisz:2010xx,Siefert:2014ela,Hogervorst:2011zw}.
\section{Technical Details} \label{draft:techdet}
The Euclidean version of the Lagrangian with counter terms is given below:
\begin{equation} \label{Lagrangian:eq}
\begin{split}
L = \frac{1}{2}(1+A) \partial_{\mu} \phi(x) \partial^{\mu} \phi(x) + \frac{1}{2} (m_{s}^{2}+B) \phi^{2}(x) + (1+\alpha) \partial_{\mu} h^{\dagger}(x) \partial^{\mu} h(x) \\ + (m_{h}^{2}+\beta) h^{\dagger}(x) h(x) + (1+a) \partial_{\mu} H^{\dagger}(x) \partial^{\mu} H(x) + (m_{H}^{2}+b) H^{\dagger}(x) H(x) \\ + (\lambda_{1}+C_{1}) \phi(x) h^{\dagger}(x) h(x) + (\lambda_{2}+C_{2}) \phi(x) H^{\dagger}(x) H(x)
\end{split}
\end{equation}
where $A$, $B$, $\alpha$, $\beta$, $a$, $b$, $C_{1}$, and $C_{2}$ are the coefficients due to the counter terms in the Lagrangian. The real singlet scalar field is represented by $\phi(x)$. $h(x)$ is designated for the SM Higgs boson while $H(x)$ represents the second Higgs boson. The resulting DSEs for field propagators are given below:
\begin{equation} \label{hdse:eq}
\begin{split}
D_{h}^{-1}(p)=(1+\alpha) p^{2} + m^{2}_{h} (1+\alpha) + 2 (1+A) (1+\alpha) (1+a) \sigma_{h} + \\ (\lambda_{1}+C_{1}) \int_{-\Lambda}^{\Lambda} \frac{d^{4}q}{(2\pi)^{4}} D_{s}(q) \Gamma_{1}(-p,q) D_{h}(q-p)
\end{split}
\end{equation}
\begin{equation} \label{Hdse:eq}
\begin{split}
D_{H}^{-1}(p)=(1+a) p^{2} + m^{2}_{H} (1+a) + 2 (1+A) (1+\alpha) (1+a) \sigma_{H} + \\ (\lambda_{2}+C_{2}) \int_{-\Lambda}^{\Lambda} \frac{d^{4}q}{(2\pi)^{4}} D_{s}(q) \Gamma_{2}(-p,q) D_{H}(q-p)
\end{split}
\end{equation}
\begin{equation} \label{sdse:eq}
\begin{split}
D_{s}^{-1}(p)=(1+A) p^{2} + m^{2}_{s} (1+A) + 2 (1+A) (1+\alpha) (1+a) \sigma_{s} + \\ (\lambda_{1}+C_{1}) \int_{-\Lambda}^{\Lambda} \frac{d^{4}q}{(2\pi)^{4}} D_{h}(q) \Gamma_{1}(q,-p) D_{h}(q-p)+ \\ (\lambda_{2}+C_{2}) \int_{-\Lambda}^{\Lambda} \frac{d^{4}q}{(2\pi)^{4}} D_{H}(q) \Gamma_{2}(q,-p) D_{H}(q-p)
\end{split}
\end{equation}
where the following definitions are used:
\begin{subequations} \label{mterms:eq}
\begin{align}
\beta = \alpha m^{2}_{h} + 2(1+A) (1+\alpha) (1+a) \sigma_{h}  \\
b = a m^{2}_{H} + 2(1+A) (1+\alpha) (1+a) \sigma_{H}  \\
B = A m^{2}_{s} + 2(1+A) (1+\alpha) (1+a) \sigma_{s}
\end{align}
\end{subequations} 
$\sigma_{h}$, $\sigma_{H}$, $\sigma_{s}$ are the terms to be determined during a computation. Due to their nature, above definitions do not impose any constraints on the equations. The definition for the two vertices during computations is given below:
\begin{subequations} \label{vers:eq}
\begin{align}
\Gamma_{1}(u,v)=(1+A) (1+\alpha) (1+a) \tilde{\Gamma}_{1}(u,v) \\
\Gamma_{2}(u,v)=(1+A) (1+\alpha) (1+a) \tilde{\Gamma}_{2}(u,v)
\end{align}
\end{subequations}
Hence, the DSEs for the three field propagators become
\begin{equation} \label{hfdse:eq}
\begin{split}
D^{-1}_{h}(p)=(1+\alpha) [\ p^{2} + \frac{m^{2}_{h,r}}{(1+\alpha)} + (\lambda_{1}+C_{1})(1+A)(1+a) \\ \int_{-\Lambda}^{\Lambda} \frac{d^{4}q}{(2\pi)^{4}} D_{s}(q) \tilde{\Gamma}_{1}(-p,q) D_{h}(q-p) ]\
\end{split}
\end{equation}
\begin{equation} \label{Hfdse:eq}
\begin{split}
D^{-1}_{H}(p)=(1+a) [\ p^{2} + m^{2}_{H} + 2 (1+A) (1+\alpha) \sigma_{H} + (\lambda_{2}+C_{2})(1+A)(1+\alpha) \\ \int_{-\Lambda}^{\Lambda} \frac{d^{4}q}{(2\pi)^{4}} D_{s}(q) \tilde{\Gamma}_{2}(-p,q) D_{H}(q-p) ]\
\end{split}
\end{equation}
\begin{equation} \label{sfdse:eq}
\begin{split}
D^{-1}_{s}(p)=(1+A) [\ p^{2} + m^{2}_{s} + 2 (1+a) (1+\alpha) \sigma_{s} + (\lambda_{1}+C_{1})(1+a)(1+\alpha) \\ \int_{-\Lambda}^{\Lambda} \frac{d^{4}q}{(2\pi)^{4}} D_{h}(q) \tilde{\Gamma}_{1}(q,-p) D_{h}(q-p) + (\lambda_{2}+C_{2})(1+a)(1+\alpha) \\ \int_{-\Lambda}^{\Lambda} \frac{d^{4}q}{(2\pi)^{4}} D_{H}(q) \tilde{\Gamma}_{2}(q,-p) D_{H}(q-p) ]\
\end{split}
\end{equation}
where in equation \ref{hfdse:eq} the renormalized squared mass for the SM Higgs is fixed at its squared physical mass. Equations \ref{hfdse:eq}-\ref{sfdse:eq} are the three DSEs from which field propagators and other quantities are numerically extracted.
\par
Lastly, the quantities $\tilde{\Gamma}_{1}(u,v)$ and $\tilde{\Gamma}_{2}(u,v)$ are fixed at $\lambda_{1}$ and $\lambda_{2}$, respectively \cite{Roberts:1994dr}. However, in the current investigation the vertices can still change depending upon the contributions from the coefficients in the counter terms.
\par
For each of the propagators, the following renormalization conditions are used.
\begin{equation} \label{hcond:eq}
D_{h}^{ij}(p)  |_{p^{2}=m^{2}_{h,r}} = \frac{\delta ^{ij}}{p^{2}+m^{2}_{h,r}} |_{p^{2}=m^{2}_{h,r}}
\end{equation}
\begin{equation} \label{Hcond:eq}
D_{H}^{ij}(p)  |_{p^{2}=m^{2}_{H}} = \frac{\delta ^{ij}}{p^{2}+m^{2}_{H}} |_{p^{2}=m^{2}_{H}}
\end{equation}
\begin{equation} \label{scond:eq}
D_{s}(p)  |_{p^{2}=m^{2}_{s}} = \frac{1}{p^{2}+m^{2}_{s}} |_{p^{2}=m^{2}_{s}}
\end{equation}
In addition, the following two conditions are imposed to numerically extract the correlation functions and the other quantities which are introduced for the counter terms.
\begin{equation} \label{hleast:eq}
\begin{split}
\int_{-\Lambda}^{\Lambda} (\ -D^{-1}_{h}(p)+ (1+\alpha) [\ p^{2} + \frac{m^{2}_{h,r}}{(1+\alpha)} + (\lambda_{1}+C_{1})(1+A)(1+a) \\ \int_{-\Lambda}^{\Lambda} \frac{d^{4}q}{(2\pi)^{4}} D_{s}(q) \tilde{\Gamma}_{1}(-p,q) D_{h}(q-p) ]\ )\ ^{2} dp =0
\end{split}
\end{equation}
\begin{equation} \label{Hleast:eq}
\begin{split}
\int_{-\Lambda}^{\Lambda} (\ -D^{-1}_{H}(p)+(1+a) [\ p^{2} + m^{2}_{H} + 2 (1+A) (1+\alpha) \sigma_{H} + \\ (\lambda_{2}+ C_{2})(1+A)(1+\alpha) \int_{-\Lambda}^{\Lambda} \frac{d^{4}q}{(2\pi)^{4}} D_{s}(q) \tilde{\Gamma}_{2}(-p,q) D_{H}(q-p) ]\ )\ ^{2} dp =0
\end{split}
\end{equation}
Equations \ref{hleast:eq} - \ref{Hleast:eq} are indeed the implementation of least square method with errors $E_{1}$ and $E_{2}$ defined below.
\begin{equation} \label{herr:eq}
\begin{split}
E_{1}= \int_{-\Lambda}^{\Lambda} (\ -D^{-1}_{h}(p)+ (1+\alpha) [\ p^{2} + \frac{m^{2}_{h,r}}{(1+\alpha)} + (\lambda_{1}+C_{1})(1+A)(1+a) \\ \int_{-\Lambda}^{\Lambda} \frac{d^{4}q}{(2\pi)^{4}} D_{s}(q) \tilde{\Gamma}_{1}(-p,q) D_{h}(q-p) ]\ )\ ^{2} dp
\end{split}
\end{equation}
\begin{equation} \label{Herr:eq}
\begin{split}
E_{2}=\int_{-\Lambda}^{\Lambda} (\ -D^{-1}_{H}(p)+(1+a) [\ p^{2} + m^{2}_{H} + 2 (1+A) (1+\alpha) \sigma_{H} + \\ (\lambda_{2}+C_{2})(1+A)(1+\alpha) \int_{-\Lambda}^{\Lambda} \frac{d^{4}q}{(2\pi)^{4}} D_{s}(q) \tilde{\Gamma}_{2}(-p,q) D_{H}(q-p) ]\ )\ ^{2} dp
\end{split}
\end{equation}
With imposition of these constraints, the problem at hand becomes that of optimization in which solutions are sought which satisfy equations \ref{hleast:eq}-\ref{Hleast:eq}.
\par
Additional conditions given below are imposed in order to ensure positivity of renormalized squared masses.
\begin{equation} \label{mass2cond:eq}
\begin{split}
m^{2}_{H,r} = (1 + a) (\ m^{2}_{H} + 2(1+A) (1+\alpha) \sigma_{H} )\ \geq 0 \\
m^{2}_{s,r} = (1 + A) (\ m^{2}_{s} + 2(1+\alpha) (1+a) \sigma_{s} )\ \geq 0
\end{split}
\end{equation}
The renormalized SM Higgs mass is fixed \footnote{The parameter $\sigma_{h}$ needs not to be calculated since the SM renormalized mass is fixed.}, as mentioned earlier.
\par
In order to suppress numerical fluctuations, the SM Higgs is expanded in the form given below:
\begin{equation} \label{hexp1:eq}
D^{ij}_{h}(p)= \delta^{ij} \frac{1}{c(p^{2}+d+f(p))}
\end{equation}
with $f(p)$ given by
\begin{equation} \label{hexp2:eq}
f(p) = \frac{\displaystyle \sum_{l=0}^{N} a_{l} p^{2l}}{\displaystyle \sum_{l=0}^{N} b_{l} p^{2l}}
\end{equation}
In equations \ref{hexp1:eq}-\ref{hexp2:eq}, $c$, $d$, $a_{l}$, and $b_{l}$ are the coefficients to be determined during a computation. A similar expansion with different coefficients is used for the second Higgs. Beside stability, these expansions are also time efficient while performing renormalization and updating the SM and the second Higgs propagators.
\par
The computation starts with $\sigma_{H}=\sigma_{s}=C_{1}=C_{2}=0$, i.e. with no contribution by the counter terms to the renormalized masses and the two Yukawa couplings. Both Higgs propagators are also rendered their respective tree level structures. For the SM Higgs in equations \ref{hexp1:eq}-\ref{hexp2:eq}, $c=1$ and $d=m^{2}_{h}$ \footnote{Here $m^{2}_{h}=m^{2}_{h,r}$ is used throughout the study.} while all the other coefficients are zero. A similar setup of coefficients is used for the second Higgs. From the renormalization conditions in equations \ref{hcond:eq} and \ref{Hcond:eq}, the terms $1+\alpha$ and $1+a$ are found. The scalar propagator assumes the values using \ref{sdse:eq} and the quantity $1+A$ is calculated by the renormalization condition \ref{scond:eq} \footnote{The scalar propagator is calculated without the term $1+A$ and then then renormalization condition sets the value of the term $1+A$.}.
\par
An iteration involves updating of the correlation functions and parameters. During an iteration, first, the $\sigma_{s}$, $\sigma_{H}$, $C_{1}$, and $C_{2}$ are updated as in the mentioned ordered. The update is performed using Newton Raphson method with the criteria imposed by the least square method in equations \ref{hleast:eq}-\ref{Hleast:eq}. The updated value is accepted only when both of the errors $E_{1}$ and $E_{2}$ reduce, see equations \ref{herr:eq}-\ref{Herr:eq}.
\par
It is followed by the SM Higgs propagator for which the coefficients in equations \ref{hexp1:eq}-\ref{hexp2:eq} are updated with the above mentioned criteria of acceptance. Upon each change, the SM Higgs propagator is calculated from equations \ref{hexp1:eq}-\ref{hexp2:eq} and renormalized using equation \ref{hcond:eq}.
\par
Lastly, the second Higgs propagator is updated using the same procedure as described above for the SM Higgs, but using equation \ref{Hcond:eq} for renormalization. Upon every change, the scalar propagator is calculated from equation \ref{sfdse:eq} and renormalized using equation \ref{scond:eq}, as mentioned earlier.
\par
A computation concludes only when there are either no further improvements in the quantities such that $E_{1}$ and $E_{2}$ are further reduced, or both of these errors are equal or below the preset value of tolerance. The value of tolerance is set at $10^{-20}$.
\par
Gauss quadrature algorithm is used for numerical integration in the DSEs. The algorithms are developed in C++ environment.
\par
The correlation functions and the parameters due to renormalization were found practically uneffected by the order of updates performed on the mentioned quantities. It is taken as the definition of uniqueness for the current investigation.
\section{Field Propagators} \label{draft:props}
\begin{figure}
\centering
\includegraphics[width=\linewidth]{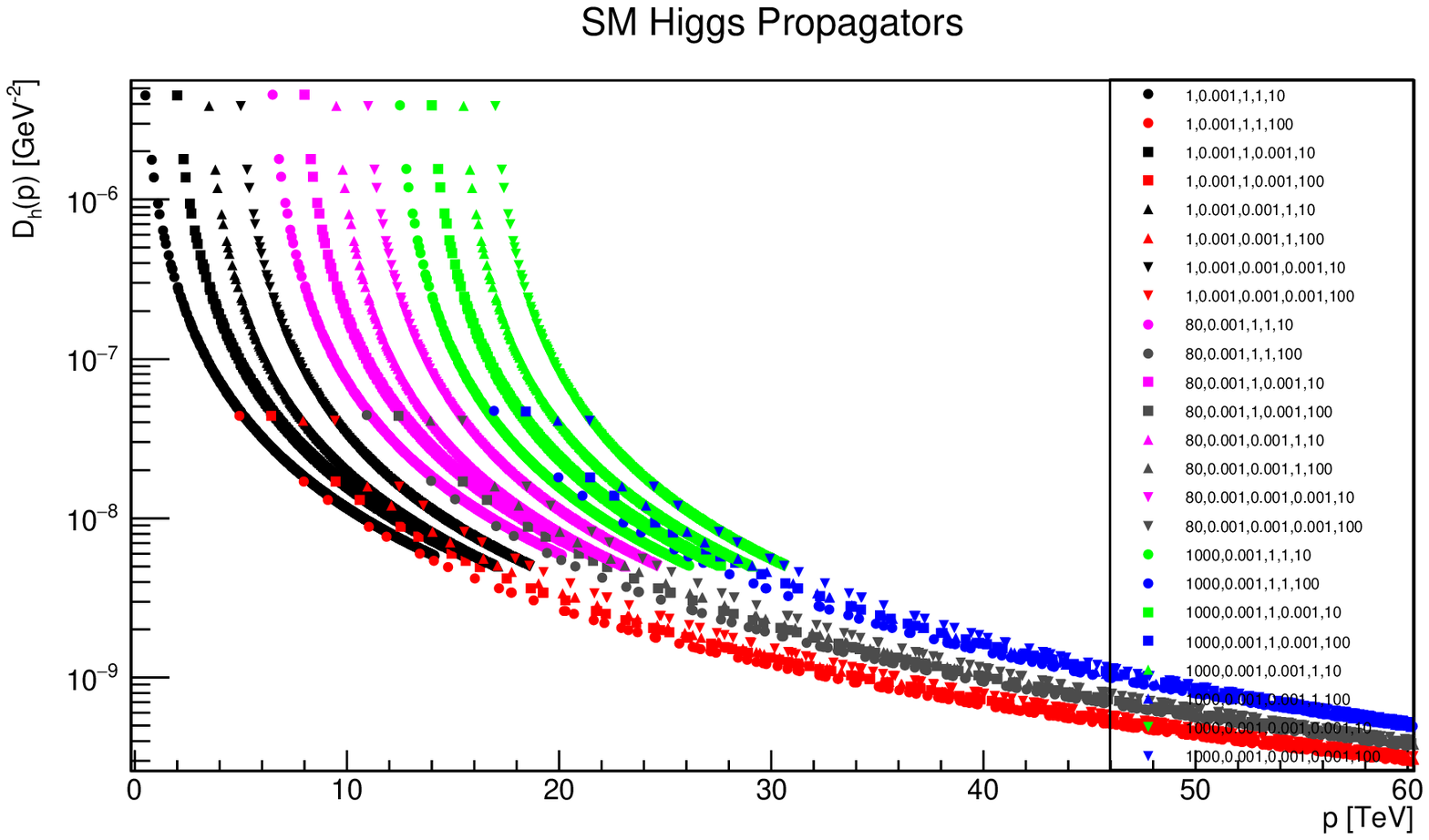}
\caption{\label{fig:h1prh21MeV} SM Higgs propagators with $m_{H}=0.001$ GeV. The parameters in the legend are given as $(m_{s},m_{H},\lambda_{1},\lambda_{2},\Lambda)$ with all the parameters but $\Lambda$ are mentioned in GeV. $\Lambda$ is given in TeV.}
\end{figure}
% \begin{figure}
% \centering
% \includegraphics[width=\linewidth]{h1prh280GeV.eps}
% \caption{\label{fig:h1prh280GeV} SM Higgs propagators with $m_{H}=80$ GeV. The parameters in the legend are given as $(m_{s},m_{H},\lambda_{1},\lambda_{2},\Lambda)$ with all the parameters, but $\Lambda$, are mentioned in GeV. $\Lambda$ is given in TeV.}
% \end{figure}
\begin{figure}
\centering
\includegraphics[width=\linewidth]{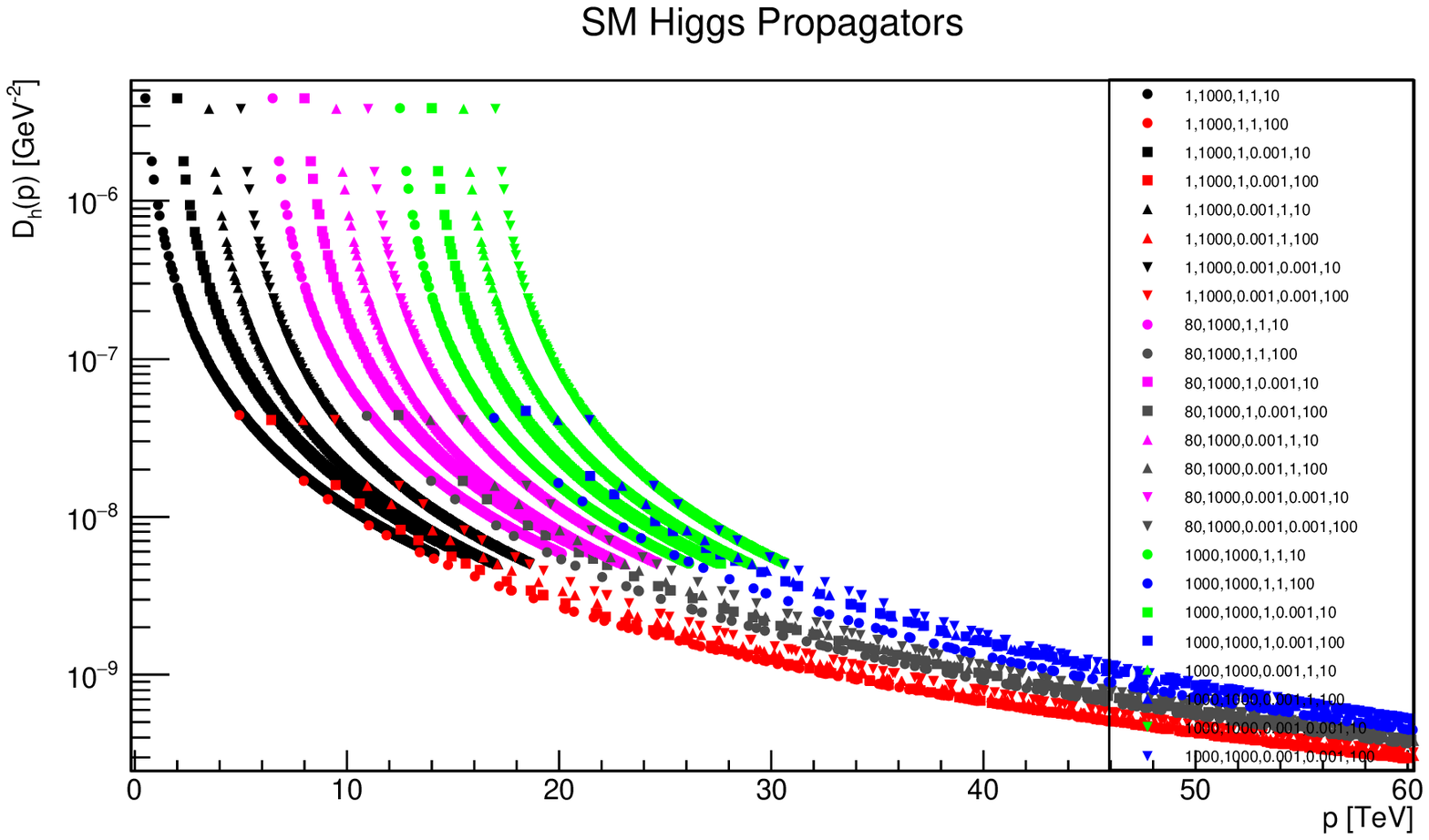}
\caption{\label{fig:h1prh21TeV} SM Higgs propagators with $m_{H}=1$ TeV. The parameters in the legend are given as $(m_{s},m_{H},\lambda_{1},\lambda_{2},\Lambda)$ with all the parameters but $\Lambda$ are mentioned in GeV. $\Lambda$ is given in TeV.}
\end{figure}
%second Higgs propagators
\begin{figure}
\centering
\includegraphics[width=\linewidth]{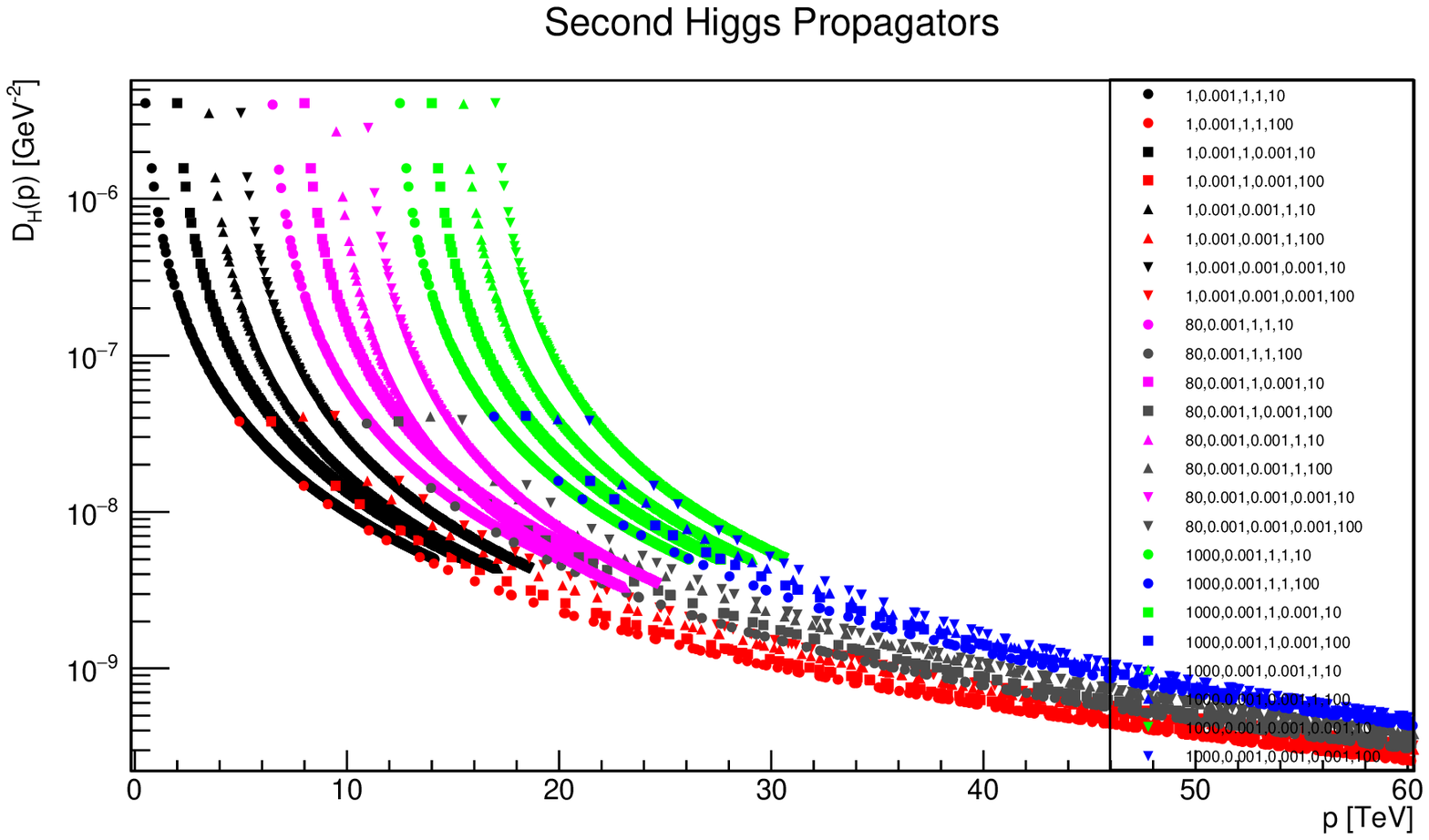}
\caption{\label{fig:h2prh20p001GeV} Second Higgs propagators with $m_{H}=0.001$ GeV. The parameters in the legend are given as $(m_{s},m_{H},\lambda_{1},\lambda_{2},\Lambda)$ with all the parameters but $\Lambda$ are mentioned in GeV. $\Lambda$ is given in TeV.}
\end{figure}
% \begin{figure}
% \centering
% \includegraphics[width=\linewidth]{h2prh280GeV.eps}
% \caption{\label{fig:h2prh280GeV} Second Higgs propagators with $m_{H}=80$ GeV. The parameters in the legend are given as $(m_{s},m_{H},\lambda_{1},\lambda_{2},\Lambda)$ with all the parameters, but $\Lambda$, are mentioned in GeV. $\Lambda$ is given in TeV.}
% \end{figure}
\begin{figure}
\centering
\includegraphics[width=\linewidth]{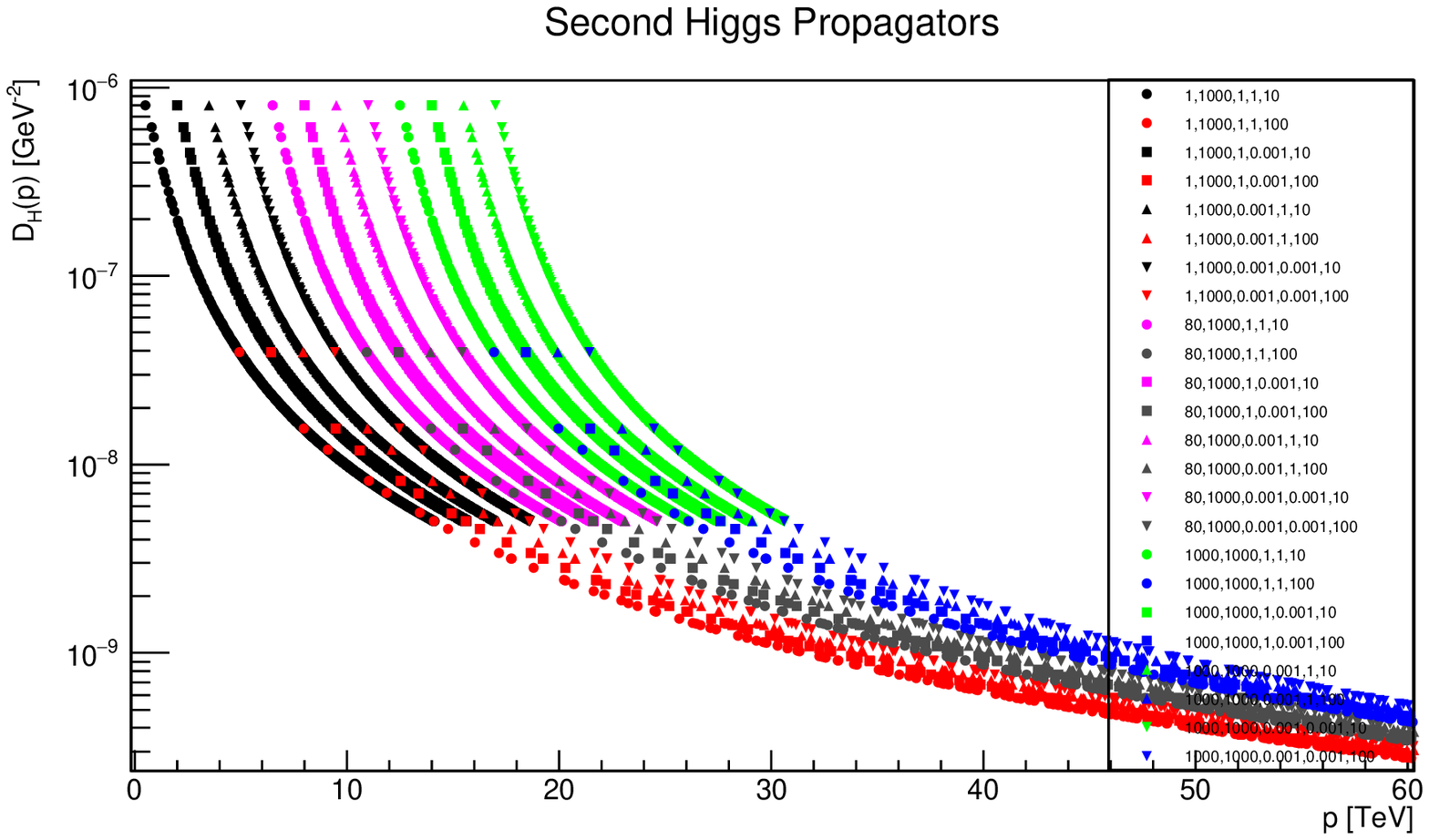}
\caption{\label{fig:h2prh21TeV} Second Higgs propagators with $m_{H}=1$ TeV. The parameters in the legend are given as $(m_{s},m_{H},\lambda_{1},\lambda_{2},\Lambda)$ with all the parameters but $\Lambda$ are mentioned in GeV. $\Lambda$ is given in TeV.}
\end{figure}
%scalar propagators
\begin{figure}
\centering
\includegraphics[width=\linewidth]{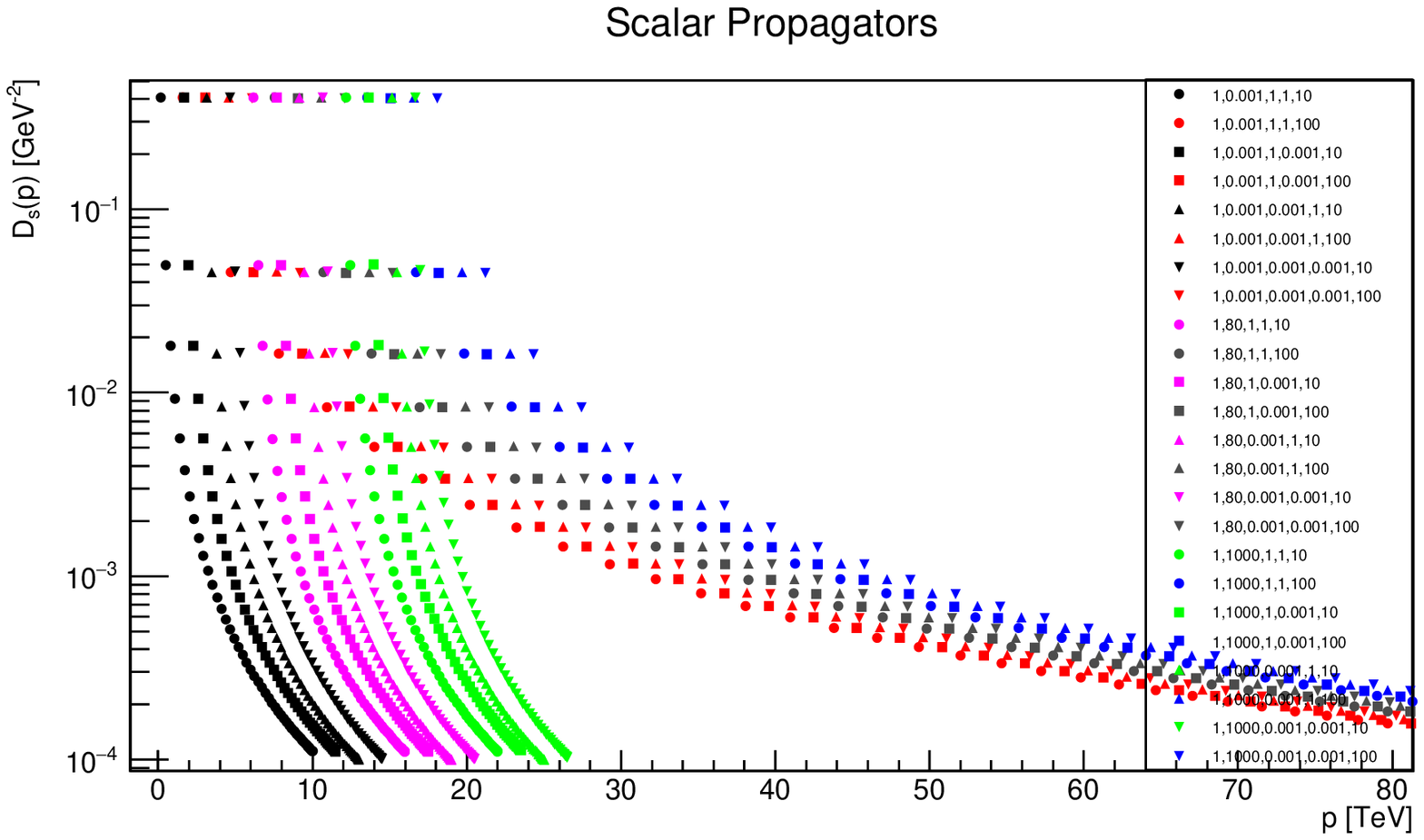}
\caption{\label{fig:spr1GeV} Scalar propagators with $m_{s}=1$ GeV. The parameters in the legend are given as $(m_{s},m_{H},\lambda_{1},\lambda_{2},\Lambda)$ with all the parameters but $\Lambda$ are mentioned in GeV. $\Lambda$ is given in TeV.}
\end{figure}
% \begin{figure}
% \centering
% \includegraphics[width=\linewidth]{spr80GeV.eps}
% \caption{\label{fig:spr80GeV} Second Higgs propagators with $m_{s}=80$ GeV. The parameters in the legend are given as $(m_{s},m_{H},\lambda_{1},\lambda_{2},\Lambda)$ with all the parameters, but $\Lambda$, are mentioned in GeV. $\Lambda$ is given in TeV.}
% \end{figure}
\begin{figure}
\centering
\includegraphics[width=\linewidth]{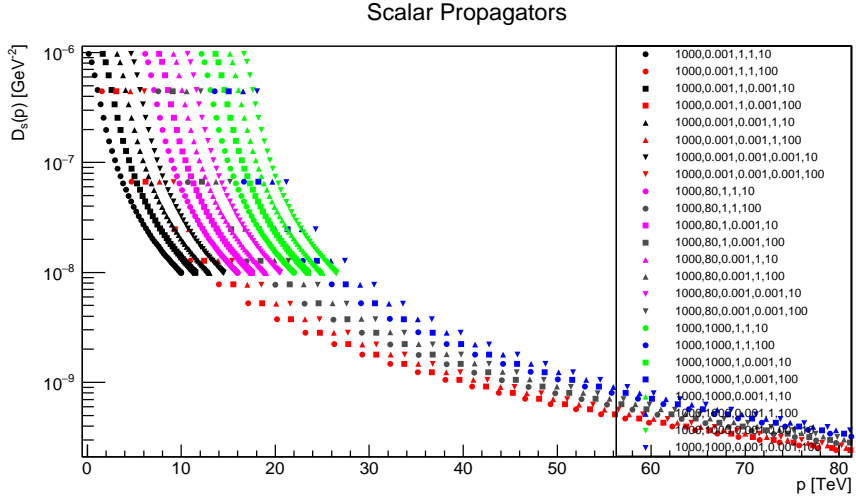}
\caption{\label{fig:spr1TeV} Scalar propagators with $m_{s}=1$ TeV. The parameters in the legend are given as $(m_{s},m_{H},\lambda_{1},\lambda_{2},\Lambda)$ with all the parameters, but $\Lambda$, are mentioned in GeV. $\Lambda$ is given in TeV.}
\end{figure}
% Field propagators are naturally effected by the choice of the vertices as they receive contributions from terms containing interaction vertices. These effects may particularly be enhanced in absence of the DSEs for the vertices. Among the propagators, the scalar singlet propagator is expected to have the largest effects since its DSE contains both of the Higgs propagators and interaction vertices in the model, see equation \ref{sfdse:eq}.
\par
The SM Higgs propagators are shown in figures \ref{fig:h1prh21MeV}-\ref{fig:h1prh21TeV}. An immediate observation is that there is no considerable cutoff effects. It implies that the propagators are mostly stable against the cutoff at or above 10 TeV. At low first coupling $\lambda_{1}$, propagators are found relatively suppressed. Hence, the propagators depend upon the first coupling to some extent while the second coupling $\lambda_{2}$ is not found to play any significant role.
\par
For the propagators of the second Higgs, despite that it shares the same symmetry as the SM Higgs field, the role of coupling and cutoff effects are relatively more pronounced, see figures \ref{fig:h2prh20p001GeV}-\ref{fig:h2prh21TeV}. One explanation may be that $m_{H,r}$ was not fixed as was the case for $m_{h,r}$. The suppression at low first coupling is relatively more pronounced. However, as the second Higgs mass is increased, the suppression is mitigated. It is not entirely unexpected since for large bare masses the defining role tend to shift towards the bare mass in a propagator, i.e. in the renormalized mass, the quantum corrections can not compete the bare mass anymore.
\par
The scalar propagators are the most effected of the three field propagators, see figures \ref{fig:spr1GeV}-\ref{fig:spr1TeV}. The cutoff effects are found significant for all of the coupling values and (bare) masses. Their severity decreases as $m_{s}$ in increased which can also be understood from the argument presented above. The propagators are found weakly depending upon the couplings, see figures \ref{fig:spr1GeV}-\ref{fig:spr1TeV}.
\par
All field propagators are found monotonically decreasing over momentum, which indicates a physical particle in the model. Furthermore, there is no pole or zero crossing \cite{Aguilar:2019uob} observed in the propagators.
\section{Renormalized Masses} \label{draft:rmass}
\begin{figure}
\centering
\includegraphics[width=\linewidth]{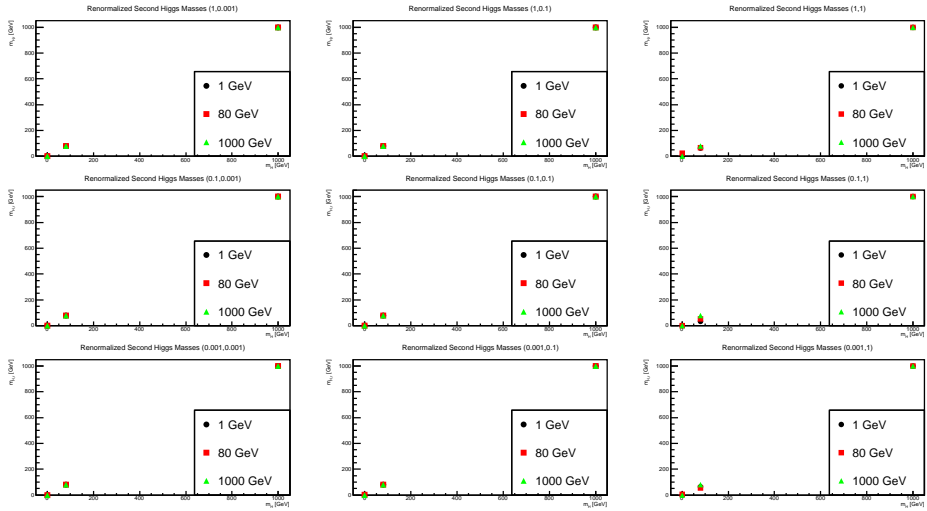}
\caption{\label{fig:h2masses10TeV} Renormalized second Higgs masses for cutoff value $\Lambda = 10$ TeV are plotted. Each diagram is plotted between second Higgs bare mass $m_{H}$ and the renormalized second Higgs mass for a particular combination of the couplings $(\lambda_{1},\lambda_{2})$, as shown in the title, for various scalar bare masses $m_{s}$, shown in the legend.}
\end{figure}
\begin{figure}
\centering
\includegraphics[width=\linewidth]{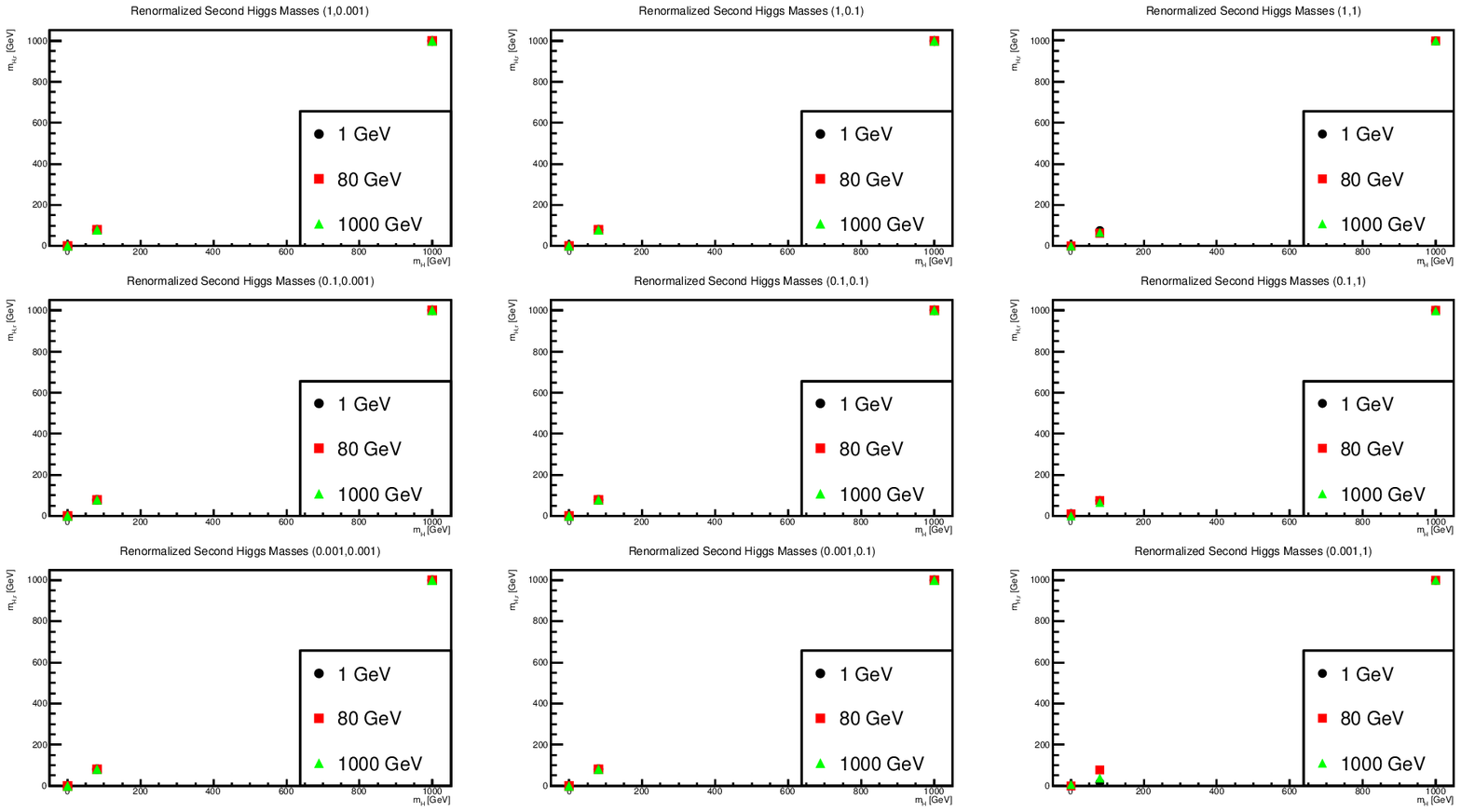}
\caption{\label{fig:h2masses100TeV} Renormalized second Higgs masses for cutoff value $\Lambda = 100$ TeV are plotted. Each diagram is plotted between second Higgs bare mass $m_{H}$ and the renormalized second Higgs mass for a particular combination of the couplings $(\lambda_{1},\lambda_{2})$, as shown in the title, for various scalar bare masses $m_{s}$, shown in the legend.}
\end{figure}
\begin{figure}
\centering
\includegraphics[width=\linewidth]{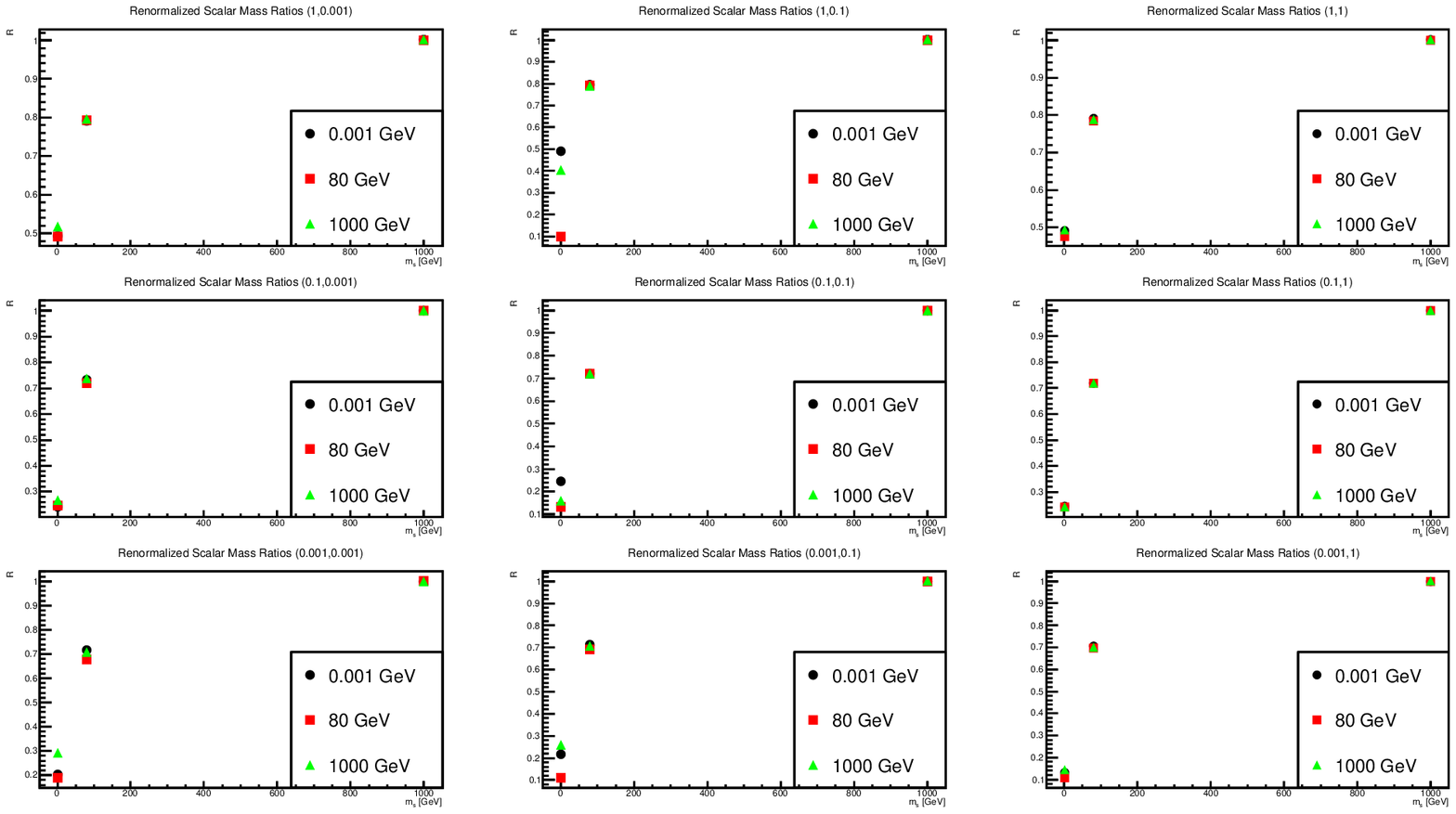}
\caption{\label{fig:sratios10TeV} Ratios $R$ between the renormalized scalar masses and their corresponding scalar bare masses $(R=\frac{m_{s,r}}{m_{s}})$ for cutoff value $\Lambda = 10$ TeV are plotted. Each diagram is plotted between scalar bare mass $m_{s}$ and the corresponding ratio for a particular combination of the couplings $(\lambda_{1},\lambda_{2})$, as shown in the title, for various second Higgs bare masses $m_{H}$, shown in the legend.}
\end{figure}
\begin{figure}
\centering
\includegraphics[width=\linewidth]{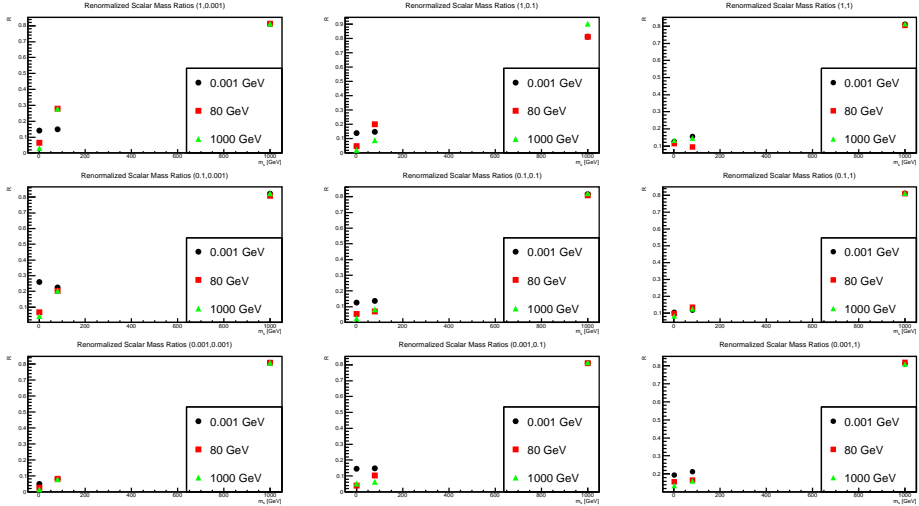}
\caption{\label{fig:sratios100TeV} Ratios $R$ between the renormalized scalar masses and their corresponding scalar bare masses $(R=\frac{m_{s,r}}{m_{s}})$ for cutoff value $\Lambda = 100$ TeV are plotted. Each diagram is plotted between scalar bare mass $m_{s}$ and the corresponding ratio for a particular combination of the couplings $(\lambda_{1},\lambda_{2})$, as shown in the title, for various second Higgs bare masses $m_{H}$, shown in the legend.}
\end{figure}
Following equation \ref{mterms:eq}, the renormalized squared masses are given by
\begin{subequations}
\begin{align}
m^{2}_{H,r} = (1+a) (\ m^{2}_{H} + 2(1+A) (1+\alpha)  \sigma_{H} )\ \\
m^{2}_{s,r} = (1+A) (\ m^{2}_{s} + 2 (1+\alpha) (1+a) \sigma_{s} )\
\end{align}
\end{subequations} \label{massdef:eq}
Stability of the second Higgs propagators against the cutoff and their weak dependence on the respective coupling can already be taken as an indication of stability in the renormalized masses.
\par
Renormalized masses for the second Higgs are given in figures \ref{fig:h2masses10TeV}-\ref{fig:h2masses100TeV}. Absence of any drastic cutoff effects or dependence on coupling(s) is evident from the figures. The second Higgs bare mass has dominant contribution to the respective renormalized mass, which only increases as the bare mass is increased to 1 TeV. In other words, the Higgs mass remains around its bare mass value in the parameter space of the model.
\par
The ratio between the renormalized and the bare masses for the scalar field are shown in figures \ref{fig:sratios10TeV}-\ref{fig:sratios100TeV}. The cutoff effects are significant on the masses which is not surprising because scalar propagators are found most effected among the field propagators.
\par
There are strong negative beyond the bare mass contributions which suppresses the renormalized masses. The suppression continues all the way to the (bare) scalar mass in TeVs. Thus, contrary to the second Higgs mass, it takes the bare mass to reach the magnitude in TeVs before it dominates in the renormalized scalar mass in terms of its contribution.
\par
A peculiar observation in the model is how the renormalized masses manifest for very low bare mass values. For both fields, the renormalized masses decrease as the corresponding bare mass decreases. It raises the question whether there is any critical coupling in the model associated with the phenomenon of dynamical mass generation. Such a behavior naturally encourages for further investigation of the model.
\section{Renormalized Couplings} \label{draft:couply}
\begin{figure}
\centering
\includegraphics[width=\linewidth]{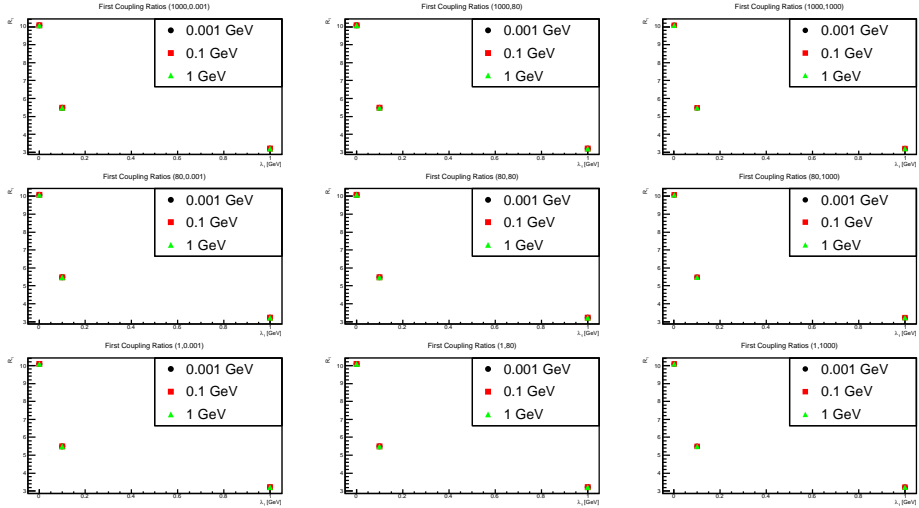}
\caption{\label{fig:lambda110TeV} Ratios $R_{1}$ between the renormalized first coupling $(\lambda_{1,r})$ and their corresponding bare coupling $(R_{1}=\frac{m_{s,r}}{m_{s}})$ for cutoff value $\Lambda = 10$ TeV are plotted. Each diagram is plotted between the first bare coupling $\lambda_{1}$ and the corresponding ratio $R_{1}$ for a particular combination of the scalar bare mass and the second Higgs bare mass $(m_{s},m_{H})$, as shown in the title, for various second bare coupling $\lambda_{2}$, shown in the legend.}
\end{figure}
\begin{figure}
\centering
\includegraphics[width=\linewidth]{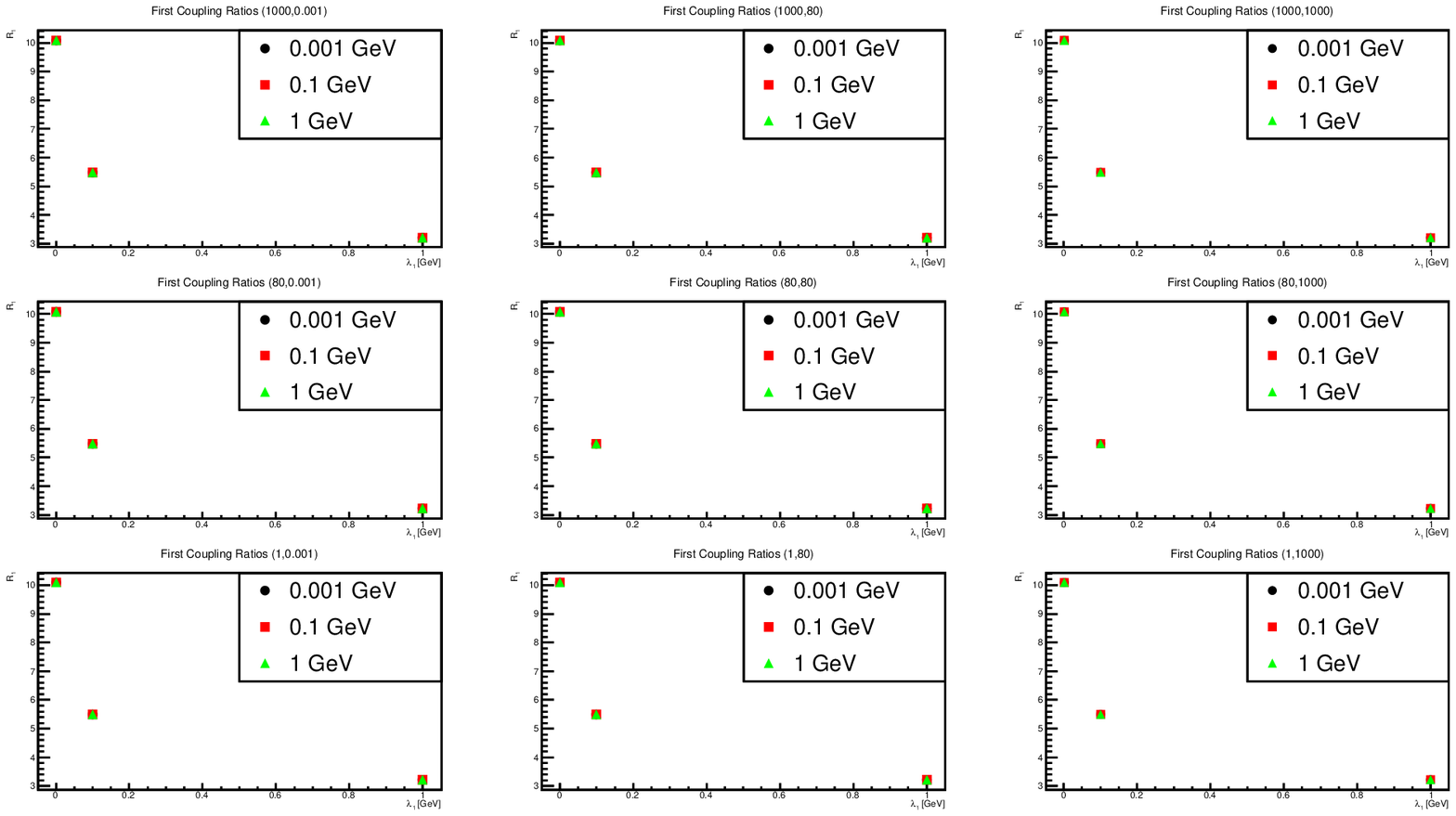}
\caption{\label{fig:lambda1100TeV} Ratios $R_{1}$ between the renormalized first coupling $(\lambda_{1,r})$ and their corresponding bare coupling $(R_{1}=\frac{\lambda_{1,r}}{\lambda_{1}})$ for cutoff value $\Lambda = 100$ TeV are plotted. Each diagram is plotted between the first bare coupling $\lambda_{1}$ and the corresponding ratio $R_{1}$ for a particular combination of the scalar bare mass and the second Higgs bare mass $(m_{s},m_{H})$, as shown in the title, for various second bare coupling $\lambda_{2}$, shown in the legend.}
\end{figure}
\begin{figure}
\centering
\includegraphics[width=\linewidth]{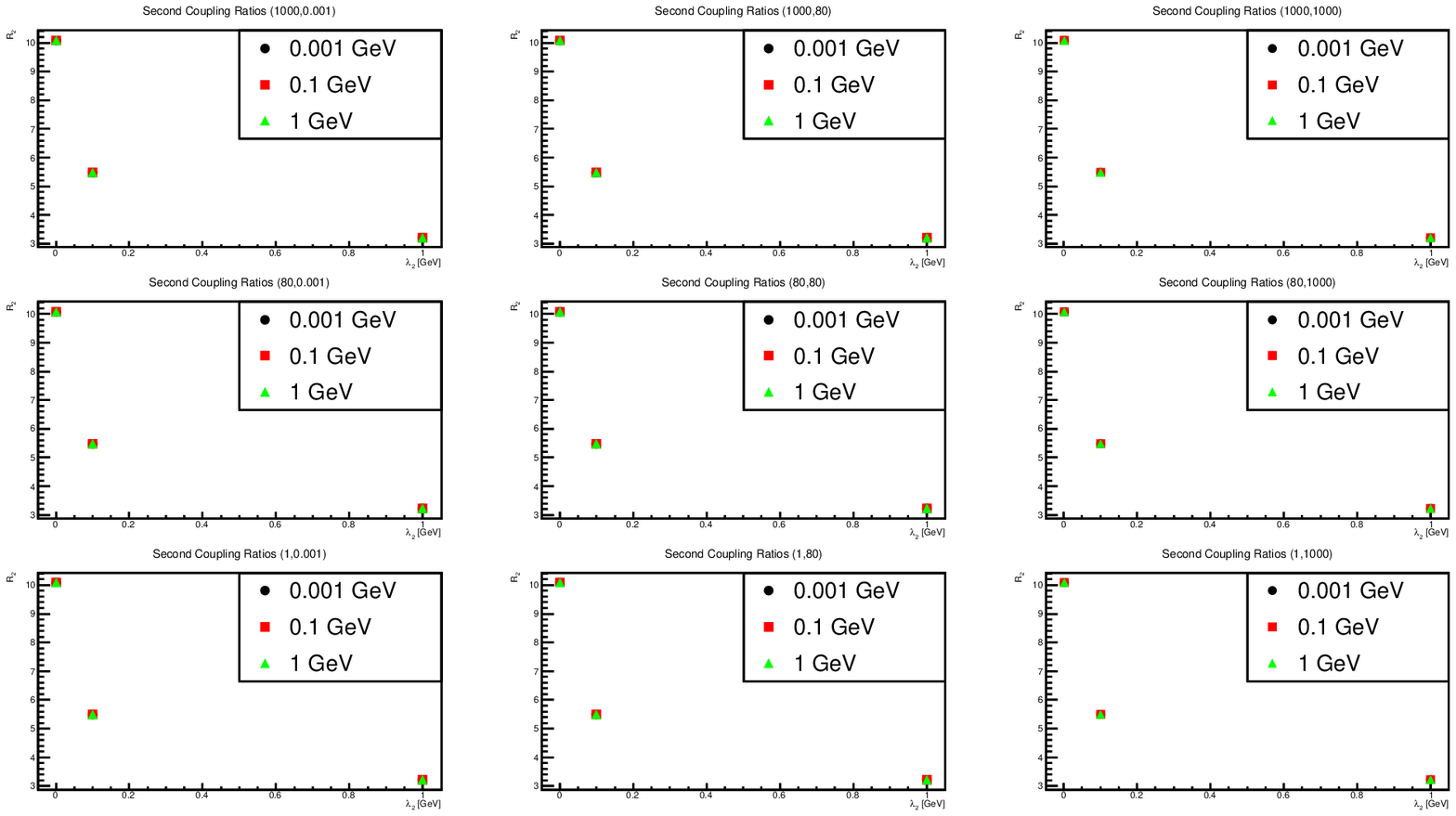}
\caption{\label{fig:lambda210TeV} Ratios $R_{2}$ between the renormalized second coupling $(\lambda_{2,r})$ and their corresponding bare coupling $(R_{2}=\frac{\lambda_{2,r}}{\lambda_{2}})$ for cutoff value $\Lambda = 10$ TeV are plotted. Each diagram is plotted between the second bare coupling $\lambda_{2}$ and the corresponding ratio $R_{2}$ for a particular combination of the scalar bare mass and the second Higgs bare mass $(m_{s},m_{H})$, as shown in the title, for various first bare coupling $\lambda_{2}$, shown in the legend.}
\end{figure}
\begin{figure}
\centering
\includegraphics[width=\linewidth]{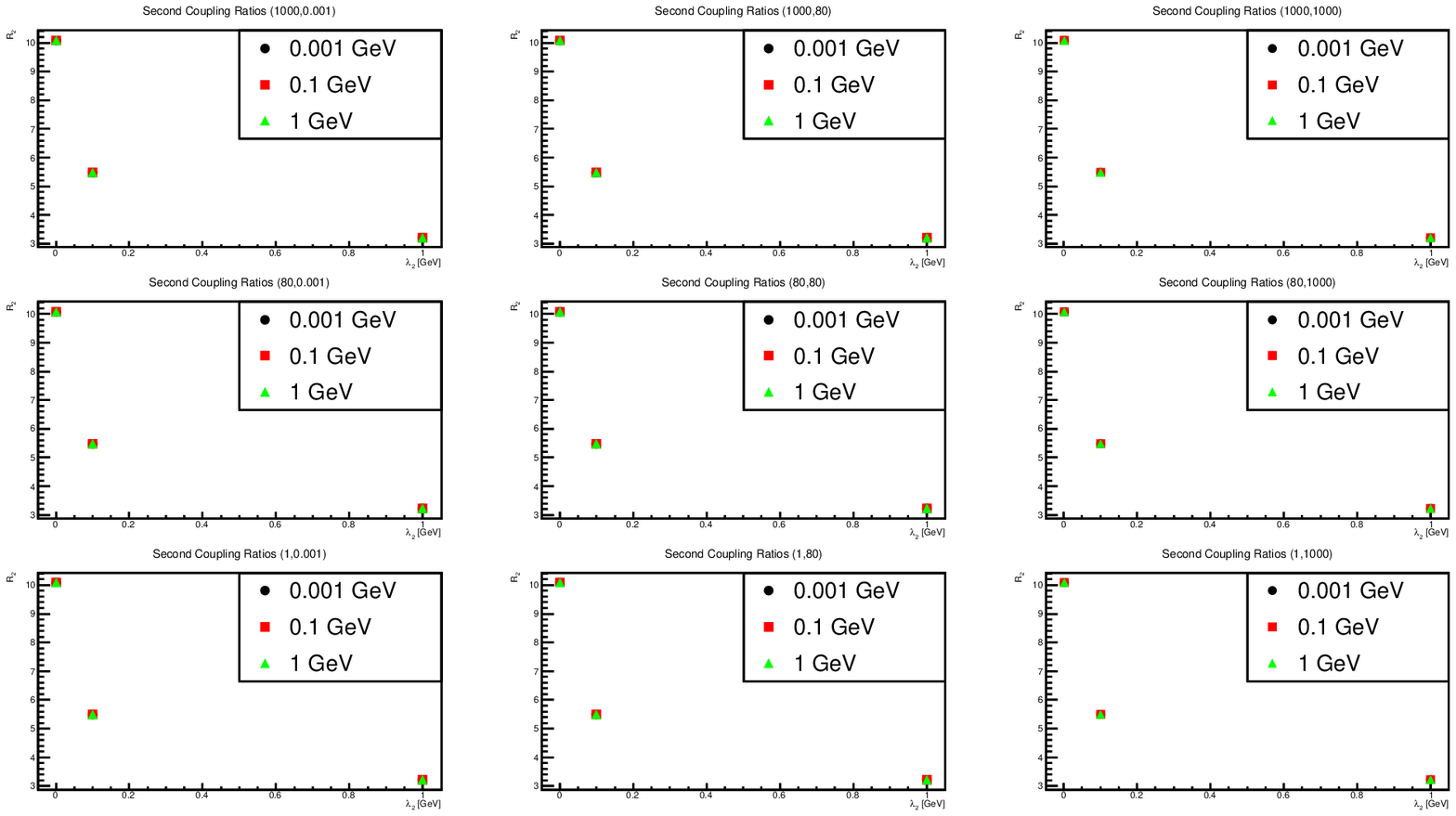}
\caption{\label{fig:lambda2100TeV} Ratios $R_{2}$ between the renormalized second coupling $(\lambda_{2,r})$ and their corresponding bare coupling $(R_{2}=\frac{\lambda_{2,r}}{\lambda_{2}})$ for cutoff value $\Lambda = 100$ TeV are plotted. Each diagram is plotted between the second bare coupling $\lambda_{2}$ and the corresponding ratio $R_{2}$ for a particular combination of the scalar bare mass and the second Higgs bare mass $(m_{s},m_{H})$, as shown in the title, for various first bare coupling $\lambda_{2}$, shown in the legend.}
\end{figure}
Following the counter terms in the Lagrangian, given in equation \ref{Lagrangian:eq}, the renormalized couplings are defined as
\begin{subequations}
\begin{align}
\lambda_{1,r} = \lambda_{1}+C_{1}  \\
\lambda_{2,r} = \lambda_{2}+C_{2}
\end{align}
\end{subequations} \label{cpldef:eq}
where  $\lambda_{1,r}$ and $\lambda_{2,r}$ are the renormalized first and second couplings, respectively.
\par
A peculiar observation in both of the renormalized couplings is that, compared with the respective bare values, they have practically the same qualitative as well as quantitative behavior. Furthermore, the results are also practically uneffected by the cutoff irrespective of the other coupling value, as was the case for both Higgs propagators for many parameters, see figures \ref{fig:lambda110TeV}-\ref{fig:lambda2100TeV}. Overall, the behavior indicates a possibility of existence of universality \cite{Crivellin:2020lzu,Azizi:2019aaf,Ciuchini:2019usw} in scalar intersection in the sense that Higgs fields of both families interact with a singlet scalar mediator in the same way. The speculation is further supported by the fact that there is practically no influence of (bare) masses. Hence, what we observe in the model is the same behavior of couplings throughout the explored region of the parameter space.
\par
The contributions in $\lambda_{i,r}$ ($i=1,2$) beyond the bare coupling values are all positive as the ratios are higher than unity in figures \ref{fig:lambda110TeV}-\ref{fig:lambda2100TeV}. It is in harmony with the suppression of the renormalized scalar masses which becomes milder for higher bare mass values. The ratios are found monotonically decreasing with a possible convergence at (bare) couplings $\lambda_{i} > 1.0$ ($i=1,2$). This is undoubtedly accessible only via non-perturbation approaches. The effects of the renormalization scheme in equations \ref{hcond:eq}-\ref{scond:eq} and the definitions of the two vertices, see equations \ref{vers:eq}, can particularly be seen here. The vertices take a particular form, and the boundary conditions are such that the scalar and second Higgs propagators are fixed at their bare mass values. Hence, as the renormalized scalar mass is lower than its respective bare values, the renormalized couplings must increase such that the scalar propagators comply with the renormalization condition. However, as the renormalized scalar mass is dominated by the respective bare masses for higher values o of the bare mass, it numerically encourages the couplings to decrease towards the bare coupling value, as is the case for both couplings.
\section{Conclusion}
Despite the simplicity of the two Higgs doublet models in comparison to richer alternatives, such as the SM, the model possesses a number of interesting features even when the interaction vertices are considered with their tree level form up to a renormalization scheme dependent constant.
\par
There are two most interesting features in the model. First, there is strong indication of universality of Yukawa coupling in the model which does not depend upon the Higgs (bare) mass. The renormalized coupling is also free of cutoff effects for all practical purposes. Second, the renormalized mass consistently decreases as the bare mass is lowered, irrespective of the couplings. It encourages for studying the phenomenon of dynamical mass generation, for example, to search for critical coupling in the model.
\par
Different quantities calculated in the model have different cutoff effects. Among the propagators, the two Higgs propagators are least effected.
\par
The model contains only two cubic vertices which are (up to the renormalization terms) kept fixed at their respective couplings. If there is a non-trivial phase structure in the parameter space for the chosen form of the vertices, it should have appeared as a deviation of qualitative behavior of the quantities calculated here. In absence of such deviations, it is concluded that the model exhibits the same physics for the chosen form of the interaction vertices. In other words, the correlation functions and the parameters describe the same phase in the parameter space of the studied model.
\section{Acknowledgments}
I am grateful to Dr. Rizwan Khalid for several discussions which were immensely helpful during the work. 
\par
This work was supported by Lahore University of Management Sciences Pakistan for development of the algorithms and numerical computations.
\bibliographystyle{plain}
\bibliography{bib}
\end{document}